# The Impact of Automation on Risk-Taking: The Role of Sense of Agency

Yang Chen, Zhijun Zhang*

*Department of Psychology and Behavioral Sciences, Zhejiang University*

*Hangzhou, China*

\* Corresponding author at: Department of Psychology and Behavioral Sciences, Zijingang Campus, Zhejiang University, No. 866 Yuhangtang Road, Hangzhou 310058, China

E-Mail address: zjzhang@zju.edu.cn (Zhijun Zhang)

Phone:13957106543

E-mail address of the authors:

Yang Chen:

12439039@zju.edu.cn

ORCID: https://orcid.org/0009-0001-4771-8277

Zhijun Zhang:

zjzhang@zju.edu.cn

ORCID: https://orcid.org/0000-0002-7194-8371

# The Impact of Automation on Risk-Taking: The Role of Sense of Agency

**Abstract**

Automation significantly alters human behavior, particularly risk-taking. Previous researches have paid limited attention to the underlying characteristics of automation and their mechanisms of influence on risk-taking. This study investigated how automation affects risk-taking and examined the role of sense of agency therein. By quantifying sense of agency through subjective ratings, this research explored the impact of automation level and reliability level on risk-taking. The results of three experiments indicated that automation reduced the level of risk-taking; higher automation level was associated with lower sense of agency and lower risk-taking, with sense of agency playing a complete mediating role; higher automation reliability was associated with higher sense of agency and higher risk-taking, with sense of agency playing a partial mediating role. The study concludes that automation influences risk-taking, such that higher automation level or lower reliability is associated with a lower likelihood of risk-taking. Sense of agency mediates the impact of automation on risk-taking, and automation level and reliability have different effects on risk-taking.

# 1. Introduction

Automation technology profoundly impacts human behavior through human-computer interaction (HCI), changing system effectiveness and user experience (Fischer, 2001). For example, automated environments notably affect risk-taking decision-making, where users may underestimate potential risks due to strong perceived control (Horswill & McKenna, 1999). Furthermore, previous researches indicated a strong correlation among automation, risk-taking, and the sense of agency (Berberian et al., 2012; Damen, 2019). Consequently, investigating how automation affects individuals' risk-taking and exploring the underlying mechanisms, such as the role of sense of agency, are of paramount importance for designing safer, more efficient automated systems and fostering harmonious human-machine coexistence. Automation is defined as the delegation of functions previously performed by humans to machine agents, typically computers. While optimizing task performance and reducing cognitive load (Wright et al., 2018), an increased level of automation can paradoxically lead to operator disengagement from the control loop and trigger some systemic risks due to diminished situational awareness or weakened fault detection capabilities (Endsley, 1997; Onnasch et al., 2014). Reliability stands as another critical factor, as its decrease tends to deteriorate operator performance (Bailey & Serbo, 2007; Ma & Kaber, 2007; Wiegmann et al., 2001; Rovira et al., 2007).

Risk-taking, is generally defined as activities offering the potential gains alongside inherent risks (Steinberg, 2010), is commonly assessed using psychometric tools like the Balloon Analog Risk Task (BART) (Lejuez et al., 2002). The influence of automation on risk-taking is complex. High-level automation assistance may induce overconfidence (Price et al., 2016), leading to an increase in risk-taking (or dangerous) activities (Brandenburg & Skottke, 2014; Navarro et al., 2019). Furthermore, previous researches on the effect of reliability present conflicting findings: some suggest higher reliability encourages more frequent automation use even with increased risk (Rodriguez et al., 2022), while others indicate low-reliability AI can also incline participants towards risk-taking (Elder et al., 2022). In conclusion, the level of automation and reliability in human-computer interaction both exert an important influence on risk-taking.

Sense of agency (SoA) is defined as the feeling of controlling one's own actions, specifically, the sensation of producing effects on the external world through one's own actions (Haggard & Chambon, 2012). It is constituted by three essential components: intention, action, and effect, with SoA significantly reinforced when effects align with anticipated goals (Barlas & Obhi, 2013). The most widely accepted comparator model posits that sense of agency originates from the matching of sensory predictions with real-time sensory feedback; enhanced consistency strengthens SoA, while discrepancies weaken it (Frith et al., 2000). SoA can be measured through direct, explicit subjective evaluations (e.g., self-report via questions like 'How much control did you have?'（Chambon & Haggard, 2012）) or indirect, implicit behavior techniques (e.g., temporal binding, physiological indicators).

Sense of agency is influenced by subjective volition, action, and causal attribution (Sato, 2009). Automation systems reduce SoA by weakening the causal link between human action intention and outcome, with this weakening effect intensifying as the level of automation increases (Berberian et al., 2012). An research by Zanatto (2021) showed a linear decrease in SoA with increased machine autonomy. In a subsequent research based on the WWW decision-making model, automation was categorized into three types of control allocation: What (content), Whether (execution), and When (timing). The findings confirmed that restricting the first two significantly reduces sense of agency, whereas timing restriction has no significant impact (Zanatto et al., 2023). The reliability of automation also affects sense of agency. Vantrepotte et al. (2023) discovered that lower system reliability correlated with higher operator SoA, inferring this was due to increased cognitive engagement.

Perceived control is negatively correlated with risk perception, a relationship that promotes risk-taking across various domains (Wang et al., 2023). However, control is a broad construct. Previous researches have found that motor-engaged control (such as actively drawing lottery balls) significantly enhances gambling risk behavior more than choice-based control, indicating that active control engagement increases risk-taking (Møller & Gregersen, 2008). Damen (2019) employed the BART paradigm, manipulating sense of agency through immediate or delayed feedback. The results revealed more risk-taking under immediate feedback. Damen thus argued that sense of agency influences risk-taking by altering risk perception. However, Karsh et al. (2021), using the same paradigm, confirmed that immediate feedback still increased risk-taking even when participants did not perceive the delay and had no conscious strategy adjustment. This suggests that sense of agency can directly influence decision-making through unconscious behavioral reinforcement pathways, independent of subjective risk assessment mechanisms.

As mentioned above, previous researches have revealed a certain relationship among automation, risk-taking, and sense of agency. Specifically, automation can influence an individual's sense of agency and risk-taking, and alterations in sense of agency can also lead to corresponding changes in risk-taking. Current researches concerning the relationship between automation and risk-taking remains relatively limited, and existing experiments often simulate specific tasks in daily life, lacking generalizability. In studies investigating the impact of automation level and reliability on risk-taking, conflicting conclusions have emerged (Rodriguez et al., 2022; Elder et al., 2022), which may be attributable to differences in experimental paradigms. In studies examining the relationship between sense of agency and risk-taking, previous researches have largely failed to quantify sense of agency, merely manipulating feedback delay to alter sense of agency, making it challenging to elucidate the underlying mechanisms of how automation influences risk-taking.

Based on the analysis above, the present study aims to investigate the characteristics of the impact of on risk-taking and the role of sense of agency therein. In accordance with this objective, this research proposes to explore two main aspects. Firstly, the study will integrate the theories and methods of sense of agency to explore the mechanism by which automation affects risk-taking. Secondly, it will investigate the impact characteristics of automation level and automation reliability on risk-taking. Thus, this research has designed three experiments. Experiment 1 serves as the foundational step, employing a variation of BART. By manipulating the fundamental automation state (automation/non-automation) through altering the balloon inflation method, it aims to initially establish whether automation influences risk-taking behavior. Building upon this foundation, Experiment 2 delves deeper by focusing on the influence of automation level. Using a task based on the variation of Experiment 1 and the WWW model (Brass & Haggard, 2008) to alter the autonomy of button presses, this experiment will concurrently measure sense of agency and risk-taking to explore the specific role of sense of agency in how automation level affects risk-taking. In parallel, Experiment 3 addresses another crucial aspect: automation reliability. Augmenting the task from Experiment 1 with a decision-aiding system, reliability is manipulated by altering the false negative rate. Similar to Experiment 2, sense of agency and risk-taking will be measured to investigate the role of sense of agency in the relationship between automation reliability and risk-taking.

## 2. Experiment 1: The Effect of Automation on Risk-Taking

This experiment employed the Balloon Analog Risk Task (BART) and referred to the methods of manipulating automation and non-automation in previous studies (Zanatto et al., 2021). Specifically, it modified the conventional BART procedure by replacing the agent responsible for button presses from a human to a computer to explore the impact of automation/non-automation on risk-taking.

## 2.1 Method

### 2.1.1 Participants

Prior to commencing the experiment, a power analysis was conducted using G*Power 3.1 (Faul et al., 2009) to estimate the required sample size. With a preset medium effect size of d = 0.5 (Cohen, 2013), a statistical power of 1-β = 0.95, and a significance level of α = 0.05, the minimum total sample size required for a t-test was determined to be 54. A total of 56 undergraduate students (24 males, 32 females) participated in this experiment. Their ages ranged from 19 to 27 years (M = 22.56, SD = 3.01). All participants had normal or corrected-to-normal vision. Upon completion of the experiment, they received monetary compensation for their participation.

### 2.1.2 Materials and design

The experimental apparatus consisted of a 24-inch monitor with a resolution of 1920×1080 and a refresh rate of 60Hz, set against a white background. The experimental program was developed using PsychoPy (Peirce, 2007). The experiment contained a variation of the Balloon Analog Risk Tasks (BART), featuring two conditions: non-automation and automation. The former was consistent with the traditional paradigm, while in the latter, the computer inflated the balloon instead of the participant. That is, this experiment employed a one-factor (automation state: non-automation vs. automation), within-subjects design.

### 2.1.3 Procedure

Participants began by completing an informed consent form. Upon entering the laboratory, they were instructed to turn on the computer audio and wear headphones in a quiet, undisturbed environment. They then read the instructions, which informed them that their reward for the balloon inflation task would be actual final cash earned. Following this, they commenced the Balloon Analog Risk Task (BART). In the non-automation condition, that is, manual inflation, the task procedure was identical to the traditional BART paradigm. Specifically, in each trial, a fixation first appeared in the center of the computer screen for 500 milliseconds. This was then replaced by a small, simulated balloon, and text appeared at the bottom of the screen stating, "Press the 'F' key to inflate the balloon once; press the 'Spacebar' key to end inflation." Participants were then instructed to begin the simulated inflation task. In detail, once the "F" key was pressed, the balloon inflated once, causing the balloon on the screen to expand correspondingly, and participants awarded the participant ¥0.01 simultaneously. Once the "Spacebar" key was pressed, the current trial finished. Each balloon had a random explosion threshold ranging from the 2nd to 45th pump. When an explosion occurred, an explosion icon appeared on the screen accompanied by an explosion sound effect, and the text "Balloon Exploded, No Earnings" was displayed, indicating that the participant received no earnings for this trial. Participants then proceeded to the next balloon trial. If participants actively chose to end inflation to secure the current earnings, the screen displayed the message "Trial Ended, Earnings: ¥X", after which they also proceeded to the next balloon trial. In the automation condition, participants initiated the first inflation by pressing the "F" key. Subsequent inflations (button pressed) were performed autonomously by the computer. To ensure that the pressing time was roughly equivalent between the non-automation and automation conditions, a pilot experiment measured the pause duration after each inflation by participants, yielding an average duration of 0.28 seconds. Therefore, each inflation occurred every 0.3 seconds after pressing the "F" key in the automation condition. During the inflation process, participants could press the "Spacebar" key at any time to end inflation for the current trial. All other details remained the same as in the non-automation inflation condition.

Before the commencement of each trial, the text "Automated Inflation" or "Manual Inflation" was displayed on the screen for two seconds to cue participants. Each participant first completed 6 practice trials to become familiar with the methods and procedures of automated and manual inflation. Following the practice session, participants proceeded to the formal experiment, each completing 60 balloon trials, divided into 6 blocks, with each block consisting of 10 trials. Tasks with different levels of automation appeared randomly within each block. Brief breaks were permitted between any two blocks. The total duration of Experiment 1 ranged from 12 to 16 minutes.

### 2.1.4 Data analyses

Due to program-related data loss, one participant was excluded from the analysis. The final dataset included 55 participants, comprising 24 males and 31 females. Python was used for data categorization and organization, and SPSS was employed for data entry and analysis. Based on prior research (Lejuez et al., 2002; Lejuez et al., 2003), the risk-taking indicators adopted in this experiment included: the average number of pumps on unexploded balloons and the number of balloons exploded. Among these, the average number of pumps on unexploded balloons served as the primary indicator of individual risk-taking, with a higher number of pumps indicating a greater propensity for risk-taking. The number of exploded balloons also reflects individual risk-taking to a certain extent, with a higher number of exploded balloons suggesting a greater propensity for risk-taking.

### 2.2 Results

The results revealed a significant main effect of automation state. The average number of pumps on unexploded balloons in the non-automation condition ($16.971 \pm 4.361$) was significantly greater than that in the automation condition ($14.805 \pm 4.093$), $t = 4.078$, $p < 0.001$, Cohen's d = 0.550 (Figure 1). The average number of exploded balloons in the non-automation condition ($24.00 \pm 83.22$) was also significantly greater than that in the automation condition ($19.55 \pm 73.68$), $t = 2.013$, $p = 0.049$, Cohen's d = 0.271.

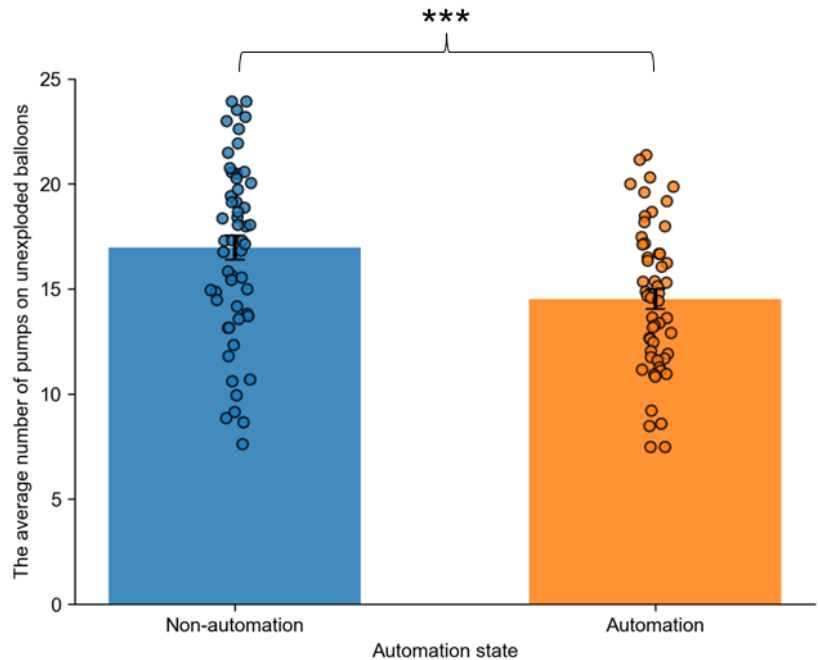

Figure 1. Average pumps on unexploded balloons: non-automation vs. automation

conditions. Error bars in all figures represent the standard error.

## 3. Experiment 2: The Effect of Automation Level on Risk-Taking: The Role of Sense of Agency

The results of Experiment 1 confirmed that individuals exhibit less risk-taking in automated contexts. However, its underlying mechanism still require further investigation. Previous researches suggest that sense of agency may be a crucial variable linking automation and risk-taking (Berberian et al., 2012; Zanatto et al., 2021; Damen, 2019). In light of this, Experiment 2 aims to further examine the impact of automation on risk-taking while monitoring changes in sense of agency to analyze how it varies with changes in automation and risk-taking. To achieve this goal, this experiment will subdivide automation into four levels based on the "What-Whether-When" model of intentional action (Brass & Haggard, 2008), and conducting a nuanced measure of sense of agency. This experiment predicts that higher levels of automation will lead to lower levels of sense of agency, and then lower levels of risk-taking.

*3.1 Method*

*3.1.1 Participants*

Prior to the experiment, a power analysis was conducted same as Experiment 1. As a result, the minimum total sample size required for ANOVA was determined to be 36. A total of 43 undergraduate students (17 males, 26 females) participated in this experiment. Their ages ranged from 19 to 29 years (M = 23.21, SD = 3.12). All participants had normal or corrected-to-normal vision and had not participated in any previous experiments of this study. Upon completion of the experiment, they received monetary compensation for their participation.

*3.1.2 Materials and design*

The experimental apparatus of Experiment 2 was the same as in Experiment 1. This experiment also employed the BART. The traditional paradigm requires multiple pumps, which is not conducive to establishing a complete sense of agency within a single trial. As mentioned earlier, Experiment 2 aims to investigate the role of sense of agency. Therefore, the inflation in all conditions of this experiment followed the "automated inflation" method of Experiment 1, meaning that inflation continued automatically after the initial press, no need to press repeatedly. Specially, it was found from a pilot study that if the time interval for automatic button presses was set at 0.3s, the connection between action and effect is easily be disrupted by lags, making it difficult for participants to establish a sense of agency. To address this, the simulated time for automatic button presses was changed to 0.1s in this experiment. Young and McCoy (2018) used a similar variation of the Balloon Analog Risk Task with a simulated time of 0.1s and found that this variation had consistent validity with the traditional manual inflation task measurement method.

This experiment also employed a one-factor, within-subjects design. The independent variable was the level of automation, which was manipulated at four levels (high automation, medium automation, low automation, and no automation) based on the "What-Whether-When" model of intentional action (Brass & Haggard, 2008), representing different task allocations between humans and machines. The specific settings were as follows: (1) High automation: Before each trial, the text "Inflation will start automatically at a random time" was presented. Subsequently, the balloon would automatically begin inflating after 2 or 4 seconds, meaning participants did not need to press any key to initiate inflation. In this condition, the individual was not assigned any part of the inflation process. (2) Medium automation: Before each trial, the text "Press a key to start inflation within 2 seconds after the text appears" was presented. Subsequently, after 2 or 4 seconds, text would reappear on the screen, instructing participants to press the "A" or "L" key to start inflation within 2 seconds, otherwise the balloon would start inflating automatically. In this condition, the task allocated to the participant was pressing a key, while the task of the machine was to specify the timing and key of the press. (3) Low automation: Before each trial, the text "Press the required letter key to start inflation" was presented. Subsequently, text would reappear on the screen, instructing participants to press the "A" or "L" key to start inflation. In this condition, the task allocated to the individual was to freely choose the timing and perform the key press, while the task of the machine was to specify the key to be pressed. (4) No automation: Before each trial, the text "Press any key to start inflation" was presented. Subsequently, participants could press any key at any time to start inflation. In this condition, the task allocated to the individual was to choose the timing and the key to be pressed, and perform the key press, and the machine was not assigned any task. The dependent variables were the same as Experiment 1 except for adding the rating of sense of agency after each balloon pump.

### 3.1.3 Procedure

Participants began by completing an informed consent form. They then read the instructions. The instructions explained the meaning of sense of agency and provided examples to help clarify the concept and the subsequent rating method. Participants then started the BART. After each round of balloon inflation, they were asked to rate their sense of agency for that inflation task on a 7-point scale adapted from the question by Chambon & Haggard (2012): "To what extent did you feel in control of the inflation?". The procedure of this experiment was generally similar to the automated inflation condition of Experiment 1. Once inflation began, subsequent inflations were performed by the computer, with an automatic inflation occurring every 0.1 seconds. Participants could press the "Spacebar" key at any time to end the inflation for the current trial. The differences among trials lay in the method of inflation and the texts appeared to cue participants before the start of each trial. To prevent a floor effect in the number of pumps, each balloon was set to explode ranging from the 10th to 64th pump. Other task details were the same as in Experiment 1.

Each participant first completed 12 practice trials to become familiar with the procedures of the different inflation methods. Following the practice session, participants proceeded to the formal experiment, completing 120 balloon trials (30 trials for each automation level), divided into 6 blocks, with each block containing 20 trials. Tasks with different levels of automation appeared randomly within each block. Brief breaks were permitted between any two blocks. The total duration of Experiment 2 ranged from 25 to 30 minutes.

### 3.1.4 Data analyses

Data were categorized and organized using Python, and the Balloon Analog Risk Task indicators were the same as in Experiment 1. The sense of agency rating for each trial was included in the analysis to explore its relationship with the level of automation and risk-taking. Finally, SPSS was used to analyze the data from all 43 participants.

### 3.2 Results

A repeated-measures ANOVA revealed a significant main effect of automation level on the average number of pumps on unexploded balloons (see Figure 2), $F(3, 126) = 6.374$, $p < 0.001$, $\eta_p^2 = 0.132$. Post-hoc tests showed that the mean number of pumps at the high automation level (20.492 ± 5.284) was significantly lower than that at the medium automation level (21.193 ± 5.882) ($p = 0.040$), and the mean number of pumps at the medium automation level was significantly lower than that at the low automation level (21.714 ± 6.127) ($p < 0.001$). No other significant differences were found. The ANOVA revealed no significant effect of automation level on the average number of exploded balloons. The feedback learning effect was still observed in the experiment, but no significant effect of automation level on outcome feedback was found.

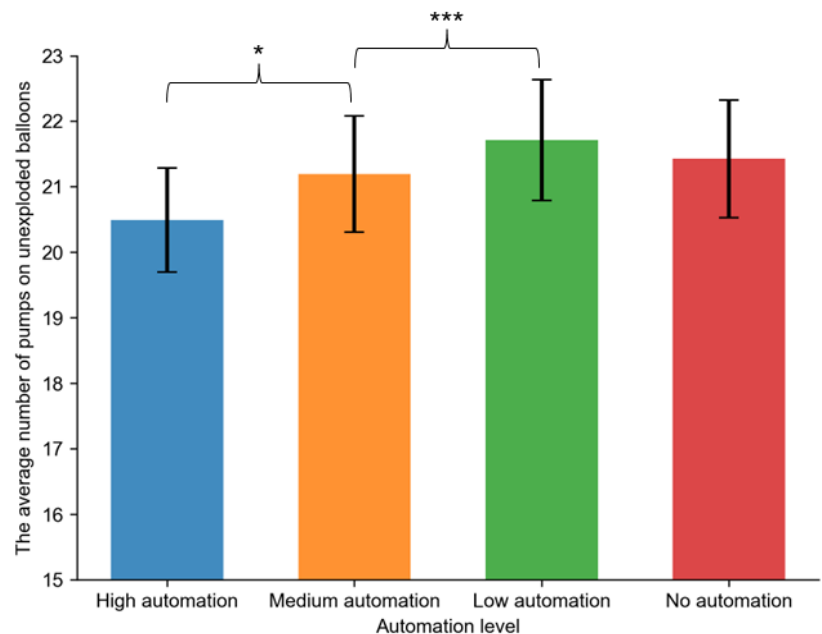

Figure 2 Average pumps on unexploded balloons by automation level

A repeated-measures ANOVA revealed a significant main effect of automation level on sense of agency (see Figure 4), $F (3, 126) = 26.789$, $p < 0.001$, $\eta_p^2 = 0.389$. Post-hoc tests showed that sense of agency was significantly lower at the high automation level (3.767 ± 1.179) compared to the medium automation level (4.099 ± 1.110) ($p = 0.002$), sense of agency was significantly lower at the medium automation level compared to the low automation level (4.581 ± 1.073) (p < 0.001), and sense of agency was significantly lower at the low automation level compared to the no automation level (5.124 ± 1.298) (*p* < 0.001). The change of sense of agency with the level of automation was generally consistent with that of the average number of pumps on unexploded balloons. Using a cutoff of 4 for sense of agency, with scores greater than 3 indicating high sense of agency and scores less than or equal to 4 indicating low sense of agency, the mean scores in sense of agency for the two groups were 5.610 ± 0.472 and 3.029 ± 0.587, respectively, showing a significant difference (*t* = 19.090, *p* < 0.001, Cohen's d = 0.845). A paired-samples t-test was used to examine the differences between the two groups, and the results showed a significant effect of sense of agency on the average number of pumps on unexploded balloons. The average number of pumps on unexploded balloons was significantly greater under high sense of agency (22.499 ± 6.237) than under low sense of agency (19.838 ± 6.237), *t* = 5.006, *p* < 0.001, Cohen's d = 0.812.

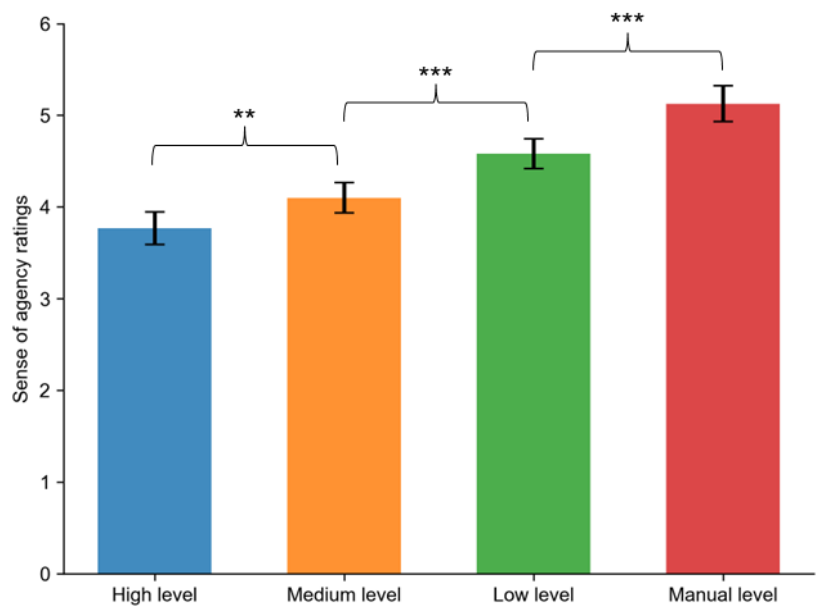

Figure 3 Average sense of agency ratings by automation level

To test the mediating role of sense of agency between automation level and risk-taking, this study used all unexploded trials from all participants as samples and employed the bootstrap method proposed by Hayes (2013), using Model 4 of the PROCESS macro in SPSS for mediation effect analysis. In the model construction, bootstrap resampling was performed 5000 times, with automation level as the independent variable (X), risk-taking (average number of pumps on unexploded balloons) as the dependent variable (Y), and sense of agency as the mediator (M). The analysis showed a significant negative indirect effect of automation level on risk-taking through sense of agency (*indirect effect* = -0.44, *S.E.* = 0.04, 95% CI = [-0.52, -0.36]). The direct effect of automation level on risk-taking was not significant ($\beta$ = -.001, *S.E.* = 0.11, 95% CI = [-0.10, 0.34]). These results indicate that sense of agency plays a complete mediating role between automation level and risk-taking. The path analysis diagram of this mediation model is shown in Figure 4.

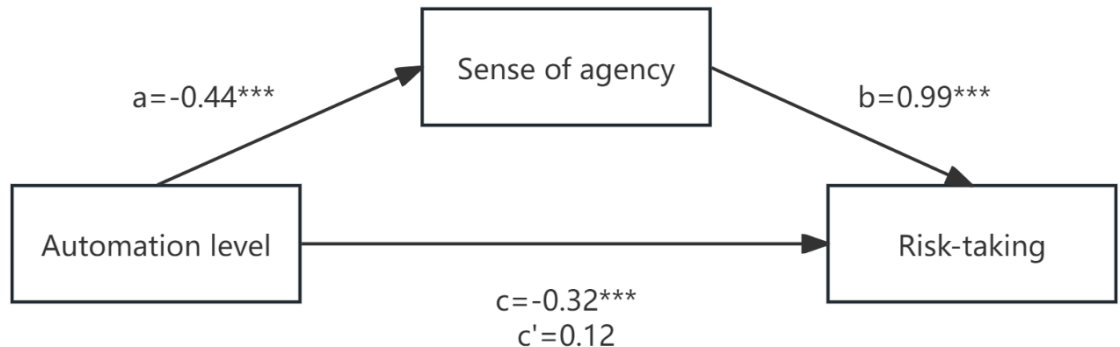

Figure 4 Mediation of sense of agency between automation level and risk-taking

## 4. Experiment 3: The Influence of Automation Reliability on Risk-Taking: The Role of Sense of Agency

The results of Experiment 2 indicated that the level of automation had a significant impact on risk-taking, with sense of agency playing a mediating role. However, the influence of automation on human behavior is multifaceted; in addition to automation level, its reliability is also a key factor. Experiment 3 will continue to employ a variation of the BART Task but will incorporate the manipulation of automation reliability into the paradigm. Specifically, an automated assistance system will be added to the modified BART. By varying the miss rate of automation, this experiment will explore the impact of automation reliability on the level of risk-taking, as well as the role of sense of agency in this relationship. This experiment predicts that higher reliability of automation will lead to a higher level of sense of agency, which in turn will be associated with a higher level of risk-taking, with sense of agency playing a mediating role.

*4.1 Method*

### 4.1.1 Participants

Prior to the experiment, a power analysis was conducted same as Experiment 1. As a result, the minimum total sample size required for ANOVA was determined to be 43. A total of 51 undergraduate students (21 males, 30 females) participated in this experiment. Their ages ranged from 18 to 29 years (M = 22.67, SD = 2.81). All participants were required to meet the same criteria as in Experiment 2, and likewise received monetary compensation.

### 4.1.2 Materials and design

The experimental apparatus used in Experiment 3 was the same as described previously. This experiment also employed the BART, with the specific procedure similar to the automated inflation condition of Experiment 1. Participants pressed the "F" key to start inflation, and subsequent inflation pumps were performed automatically by the computer at a rate of one pump every 0.1 seconds. Participants could press the "Spacebar" key at any time to end the inflation for the current trial. Additionally, Experiment 3 incorporated an automated decision support system that provided warnings with varying degrees of reliability regarding the explosion of the balloon.

Experiment 3 also used a one-factor, within-subjects design. The independent variable was the reliability of automation, which was manipulated at three levels (high reliability, medium reliability, and low reliability) based on the miss rate of the decision support system. Specifically, in the automated inflation condition, the maximum number of pumps for each balloon varied, so participants could not accurately predict the time of explosion. The role of the decision support system was to provide a warning at a random time point ranging from 45% to 55% of the total inflation capacity of the balloon. This warning consisted of the red characters "warning" appearing above the balloon for 100ms, as shown in Figure 5. The reliability of automation was manipulated by adjusting the miss rate of the decision support system (warning information): (1) High reliability: there was an 80% chance of a warning appearing in this trial; (2) Medium reliability: there was a 50% chance of a warning appearing in this trial; (3) Low reliability: there was a 20% chance of a warning appearing in this trial. Before the start of each trial, the reliability level for that trial was displayed on the screen. The dependent variables were the same as in Experiment 2.

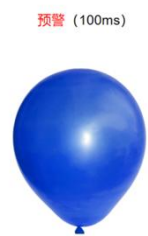

Figure 5 Location and duration of the "Warning" appearance

*4.1.3 Procedure*

The procedure of this experiment was similar to that of Experiment 2, with the exception that the balloon would explode randomly ranging from the 20th to 80th pump. All other details were the same as in Experiment 2.

Each participant first completed 3 practice trials where the warning appeared with 100% reliability to become familiar with the principle and procedure of the warning. Following the practice session, participants proceeded to the formal experiment, completing 90 trials of the BART (30 trials for each level of automation reliability), divided into 6 blocks with 15 trials per block. Tasks with different reliability levels appeared randomly within each block. Brief breaks were permitted between any two blocks. The total duration of Experiment 3 was approximately 20 to 25 minutes.

### 4.1.4 Data analyses

The data processing and statistical analyses for this experiment were identical to those used in Experiment 2. No participant data showed any abnormalities, meaning all 51 samples were included in the final statistics and analyses.

## 4.2 Results

A repeated-measures ANOVA revealed a significant main effect of automation reliability level on the average number of pumps on unexploded balloons (see Figure 6), $F$ (2, 100) = 47.360, $p < 0.001$, $\eta_p^2 = 0.538$. Post-hoc tests showed that the mean number of pumps at the low reliability level (32.450 ± 6.055) was significantly lower than that at the medium reliability level (34.611 ± 5.671) $(p < 0.001)$, and the latter was significantly lower than that at the high reliability level (37.409 ± 5.574) ($p < 0.001$). The ANOVA revealed no significant effect of automation reliability on the average number of exploded balloons. The feedback learning effect was still observed in the results of this experiment, but no significant effect of automation reliability on outcome feedback was found.

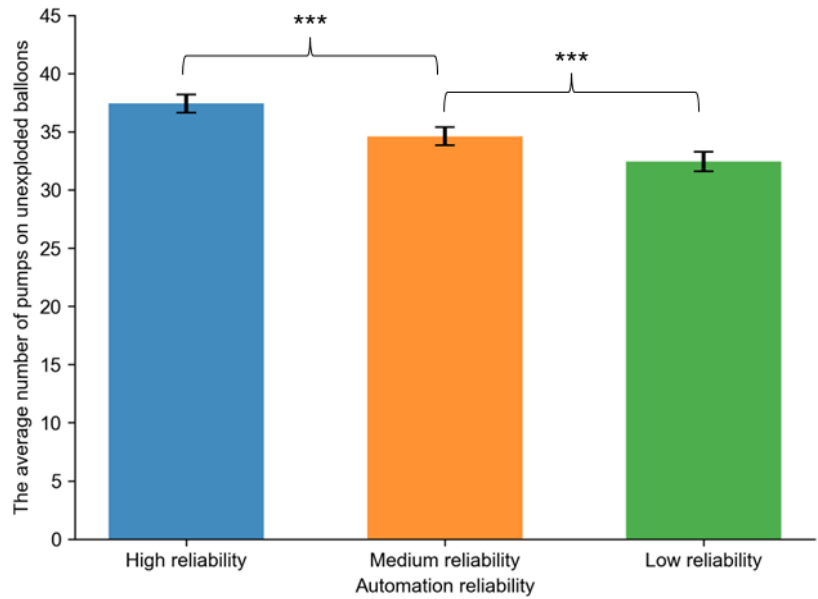

Figure 6 Average pumps on unexploded balloons by automation reliability

A repeated-measures ANOVA showed a significant main effect of reliability level on sense of agency, $F$ (2, 100) = 23.240, $p < 0.001$, $\eta_p^2$ = 0.317 (see Figure 7). Post-hoc tests showed that sense of agency was significantly lower at the low reliability level (3.455 ± 0.832) compared to the medium reliability level (3.721 ± 0.779) ($p < 0.001$), and the latter was significantly lower than that at the high reliability level (3.913 ± 0.855) ($p = 0.004$). Similar to Experiment 2, using a cutoff of 4 for sense of agency, all participants were divided into two groups. As a result, the mean scores in sense of agency for two groups were 5.216 ± 0.810 and 2.830 ± 0.500, respectively, showing a significant difference ($t$ = 18.505, $p < 0.001$, Cohen's d = 0.921). A paired-samples t-test was used to examine the differences of the average number of pumps on unexploded balloons between two groups, and it was also found a significant effect. The average number of pumps on unexploded balloons was significantly greater under high sense of agency (38.785 ± 10.069) than under low sense of agency (31.9999 ± 5.208), $t$ = 5.581, $p <$ 0.001, Cohen's d = 8.683.

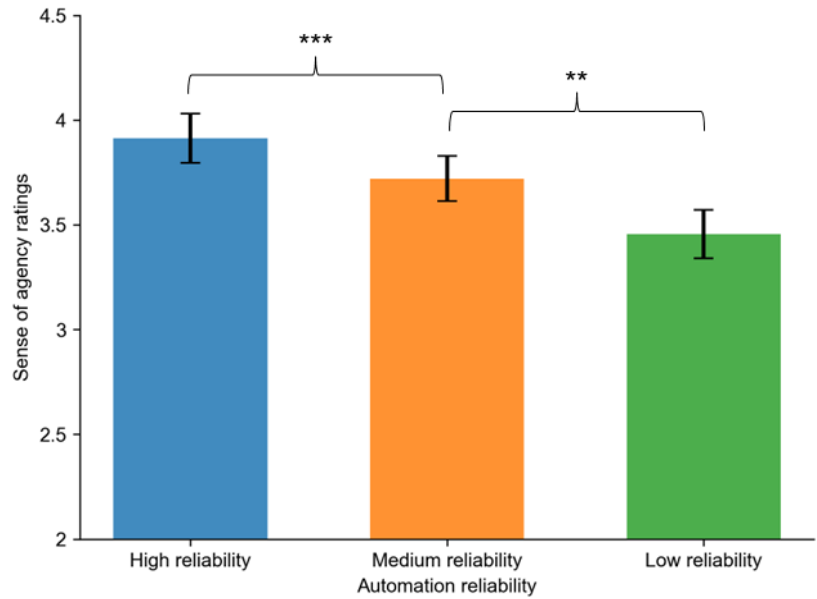

Figure 7 Average sense of agency ratings by automation reliability

To test the mediating role of sense of agency between automation reliability and risk-taking, this study used all unexploded trials from all participants as samples and employed the same method as in Experiment 2 for mediation effect analysis. That is, automation reliability was used as the independent variable (X), risk-taking (average number of pumps on unexploded balloons) as the dependent variable (Y), and sense of agency as the mediator (M). The analysis showed a significant positive indirect effect of automation reliability on risk-taking through sense of agency (*indirect effect* = 0.55, *S.E.* = 0.07, 95% CI = [0.42, 0.70]). The direct effect of automation reliability on risk-taking was also significant ($\beta$ = 2.14, *S.E.* = 0.21, 95% CI = [1.73, 2.56]). These results indicate that sense of agency plays a partial mediating role between automation reliability and risk-taking. The path analysis diagram of this mediation model is shown in Figure 8.

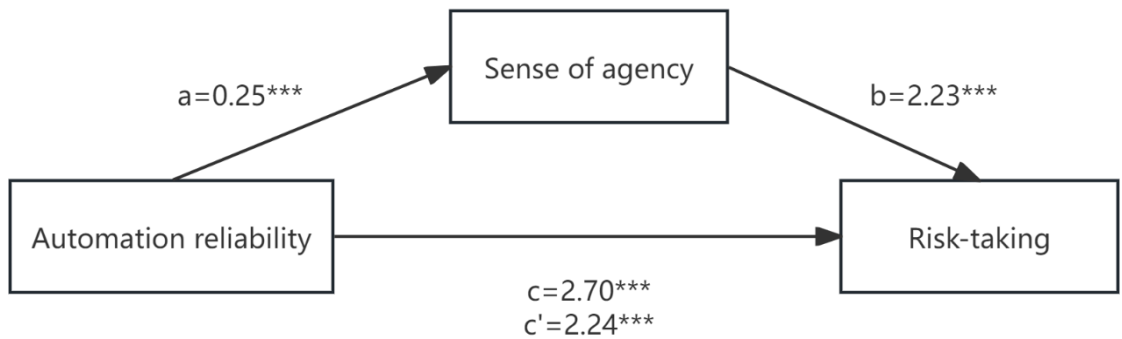

Figure 8 Mediation model of sense of agency between automation reliability and risk-taking

## 5. General Discussion

### 5.1 Impact of Automation on Risk-Taking

The results of this study confirmed the impact of automation on risk-taking, and further revealed that both the level and reliability of automation influence risk-taking, which was consistent with the expected outcomes. There are some shortcomings in the existing researches that directly examining the relationship between automation and risk-taking. They often employ experimental paradigms that simulate real-world scenarios, using behavioral performance as an indicator（Price et al., 2016）. For instance, Brandenburg and Skottke (2014) employed a driving simulator to model automated driving contexts, assessing participants' risk-taking by analyzing the distance to the lead vehicle and their speed. Elder et al. (2022) utilized a gambling task involving predicting basketball game outcomes, evaluating participants' risk-taking propensity by analyzing response bias. However, due to significant variations and complexities across different experimental settings, it is difficult to draw a universal conclusion and to apply these findings to the development of relevant theories and models. This present study modified the BART by manipulating behavioral components to manipulate automation. This paradigm reduced the interference of extraneous variables, allowing the research conclusions to be more robustly generalized to other contexts.

Some studies have already examined the underlying reasons. Stuck et al. (2021) found that the higher an individual's trust in automation, the lower their perception of risk, leading to more risk-taking; in other words, rust, by influencing risk perception, impacts individual behavior. In the context of interacting with autonomous vehicles, overtrust caused drivers to fail to take over control from automated systems in a timely manner, which is essentially a form of risk-taking (Gremillion et al., 2016). Cognitive load is also an important factor in human-automation interaction. For example, Ebadi et al. (2021) suggested that autonomous driving reduces drivers' cognitive load, leading to a decrease in situational awareness, thereby lowering risk perception and resulting in risk-taking behavior. Changes in cognitive load during the lane keeping system can also lead to individual risk compensation behavior (Miller & Boyle, 2018). However, human-automation trust, risk perception, and cognitive load are all influenced by individual factors, and their structures are relatively complex, preventing them from serving as objective and stable predictive mechanisms (Kraus et al., 2020; Reniers et al., 2016; Grassmann et al., 2016). Therefore, this study introduced the concept of sense of agency into the relationship between automation and risk-taking. Sense of agency belongs to a variable of individual's motor-perceptual level (Beck et al., 2017), is less affected by individual and social factors, and is more stable. Starting from the influencing factors of sense of agency (e.g., delay), we can also get more insights for changes in risk-taking behavior in real-world scenarios.

*5.2 Role of Sense of Agency in Automation and Risk-Taking*

Experiments 2 and 3 of this study demonstrated that automation influences sense of agency, which in turn affects risk-taking. Sense of agency plays a mediating role in the relationship between automation and risk-taking, and it is one of the underlying psychological mechanisms explaining and predicting the changes of both. More specifically, this study obtained continuous, subjective intensity data of participants' sense of agency by quantifying SoA in each task, thereby achieving a more accurate understanding of its relationship between the two.

In previous research on sense of agency and risk-taking behavior, the independent variable was manipulated at a categorical level without quantification (Damen, 2019; Karsh et al., 2021). Benefiting from the quantification of sense of agency, this study deepens the understanding of SoA as an active and influential psychological state, which can serve as an internal reward that promotes many aspects of behavior (Karsh et al., 2016). This positive control feedback is beneficial for individuals, thereby altering the characteristics and choices of individual behavior (Karsh & Eitam, 2015). The inherent functionality of sense of agency highlights its critical role in adaptive human behavior, potentially also playing an important role in the domain of risk-taking.

Building upon Zanatto's (2023) manipulation of human-machine decision-making freedom allocation in sense of agency research, Experiment 2 explored the relationship among automation level, sense of agency, and risk-taking behavior. It replicated Zanatto's effect, further validating the impact of automation level on sense of agency, as evidenced by significant changes in SoA ratings across different automation levels. Building on this, the study further expanded by discovering the mediating mechanism of sense of agency between automation level and risk-taking behavior. Furthermore, inspired by Vantrepotte's (2023) research, Experiment 3, by manipulating the reliability level of automation, also observed that SoA significantly changed with the level of reliability, along with the mediating role of SoA. In summary, this study investigated the causal relationship among automation, SoA, and risk-taking behavior, as well as the mediating role of SoA, which can provide beneficial insights for future research on SoA and risk-taking behavior.

*5.3 Influence Mechanisms of Automation Level and Reliability on Risk-Taking*

Experiments 2 and 3 found that the automation level and reliability level of human-machine systems have significant impacts on risk-taking behavior, but these impacts show opposite trends. These results suggest that automation level and reliability level may influence risk-taking behavior through different mechanisms, as sense of agency consistently mediated the relationship between the two variables in both experiments, allowing the influence mechanisms to be explained by different characteristics of sense of agency.

Based on the "intention-action-effect" model of sense of agency (Barlas & Obhi, 2013), some explanations and conjectures can be made about the paths through which automation characteristics influence risk-taking. Specifically, Experiment 2 found that as the automation level increased, sense of agency decreased, and the level of risk-taking behavior decreased. The manipulation of automation level in Experiment 2 was based on the WWW model of sense of agency, which is founded on intentional action decisions, focusing on the intention-action elements within sense of agency; therefore, the conclusions of Experiment 2 demonstrate that automation level influenced the intention-action process within sense of agency. Previous researches have also found that the introduction of automation impacts the individual's intention, and when the system behaves more intentionally, the individual's sense of agency tends to decrease more significantly (Berberian et al., 2012; Ciardo et al., 2019). In the process of increasing automation levels, the 'human-out-of-the-loop' problem becomes more severe, leading to a decrease in intention level and consequently a reduction in sense of agency, which results in individuals adopting more conservative strategies in risk-taking decisions.

Experiment 3 found that as automation reliability increased, sense of agency rose, and the level of risk-taking behavior increased. However, previous researches observed that as the automation reliability level decreased, operators' sense of agency increased (Vantrepotte et al., 2023). Unlike their study which adjusted reliability by altering the false alarm rate of the automation, this experiment adjusted reliability by changing the false negative rate (miss rate) of the automation , suggesting that manipulating reliability through misses and false alarms has different effects on sense of agency. This can be inferred based on the findings of Chancey et al. (2015) the different impacts of false alarms and misses on compliance and reliance: in situations with high false alarm rates, individuals reduce compliance, requiring cognitive resources to compensate for or identify errors; whereas in situations with low false negative rates, individuals increase their reliance on the system, enabling operators to form precise internal predictions. This result can be explained by the comparator model of sense of agency: when the system provides reliable warning information, operators can form a more accurate and trustworthy prediction based on this information and effectively execute risk management actions, thereby leading to a high match between prediction and actual results, which in turn enhances individual sense of agency. This process precisely corresponds to the action-effect process.

In summary, automation level influences the agency of individual intentions, while automation reliability affects the predictability of action effects, leading to a separation in their mechanisms of action on sense of agency and risk-taking behavior. The influence mechanisms of automation level and reliability are shown in Figure 9.

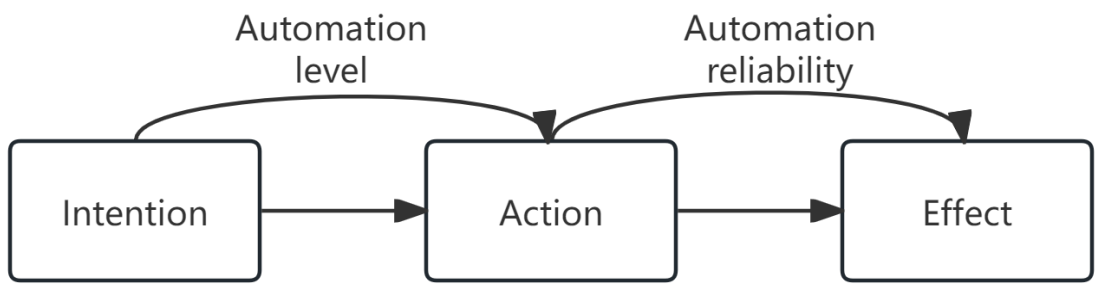

Figure 9 Influence mechanisms of automation characteristics on risk-taking

## 6. Conclusion

This study employed the Balloon Analog Risk Task to investigate the impact of automation on risk-taking and the role of sense of agency therein. Based on the three experiments, this study yielded several key findings. Automation influences risk-taking, with the level of risk-taking decreasing under automation. Both the level and reliability of automation affect risk-taking: higher automation level is associated with lower risk-taking, while higher reliability level is associated with higher risk-taking. Sense of agency plays a mediating role in the impact of automation on risk-taking. Automation influences sense of agency, and a higher level of sense of agency is associated with a higher level of risk-taking. Automation level and automation reliability influence risk-taking through different mechanisms.

**Acknowledgements**

We would like to thank all the volunteers who participated in this study.

**Declaration of interest statement**

No potential conflict of interest was reported by the author(s)

**Figures**

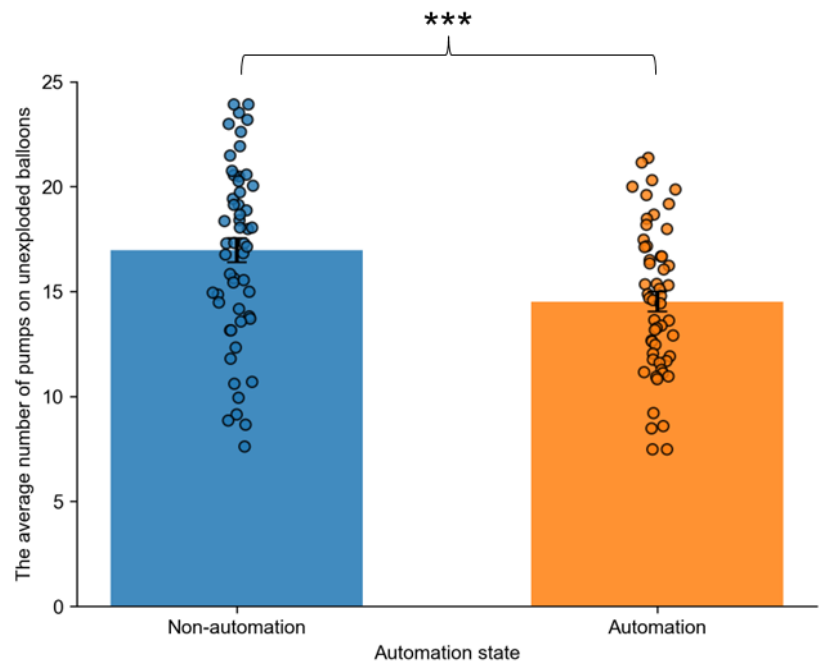

Figure 1

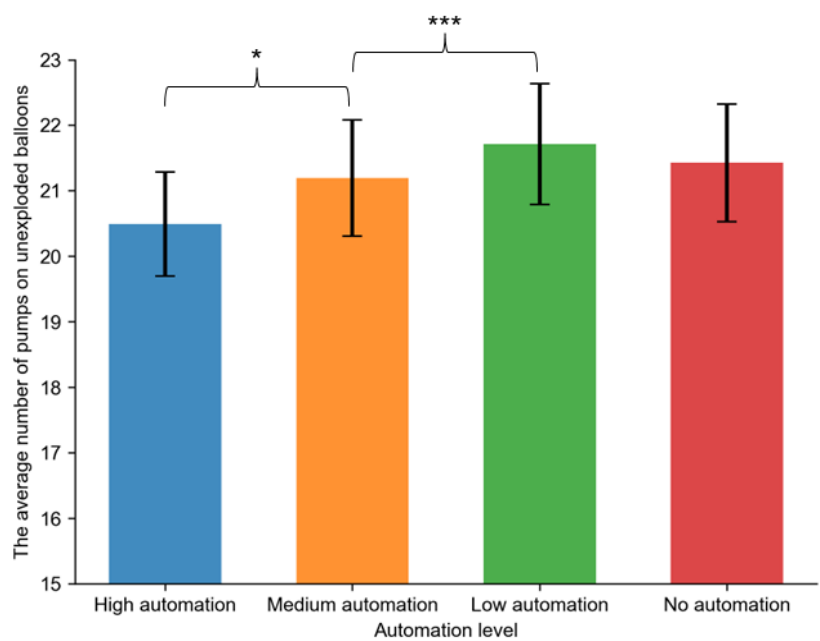

Figure 2

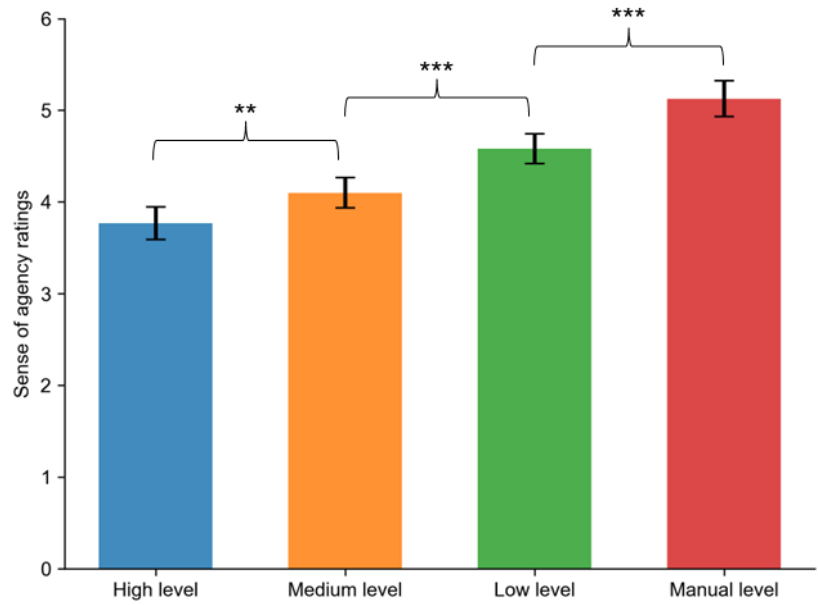

Figure 3

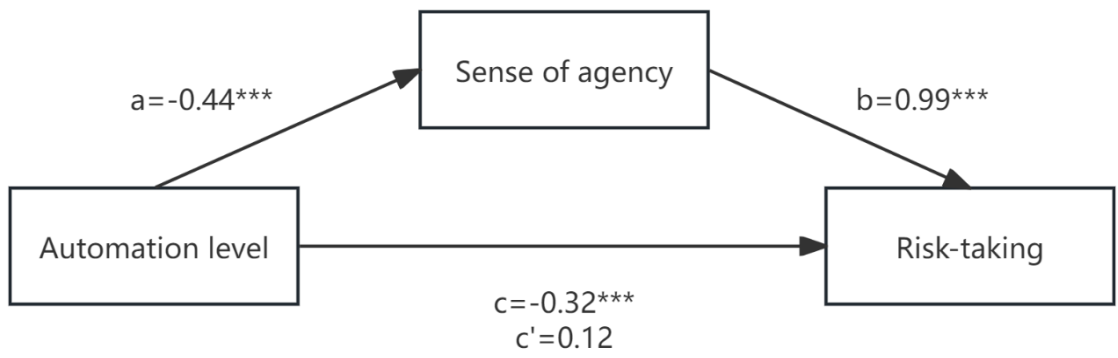

Figure 4

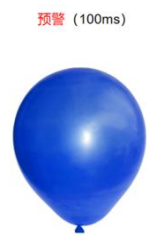

Figure 5

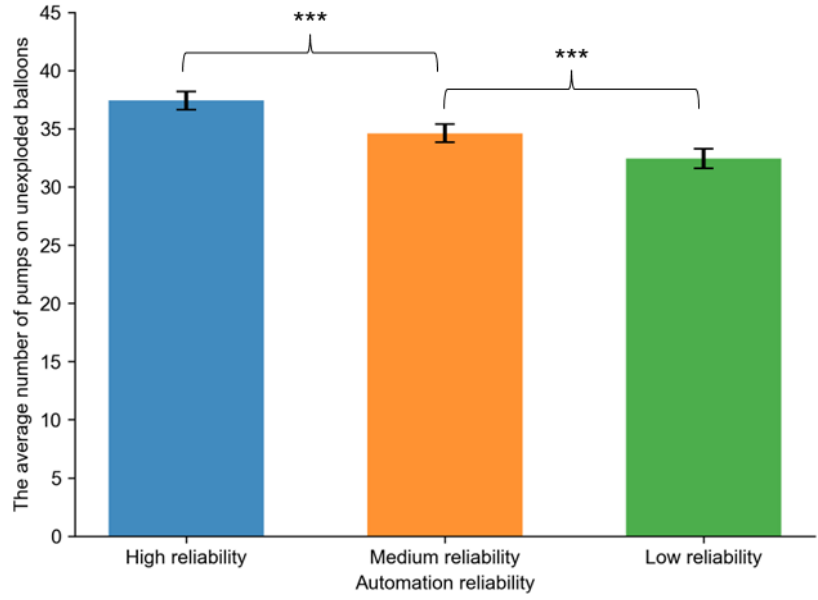

Figure 6

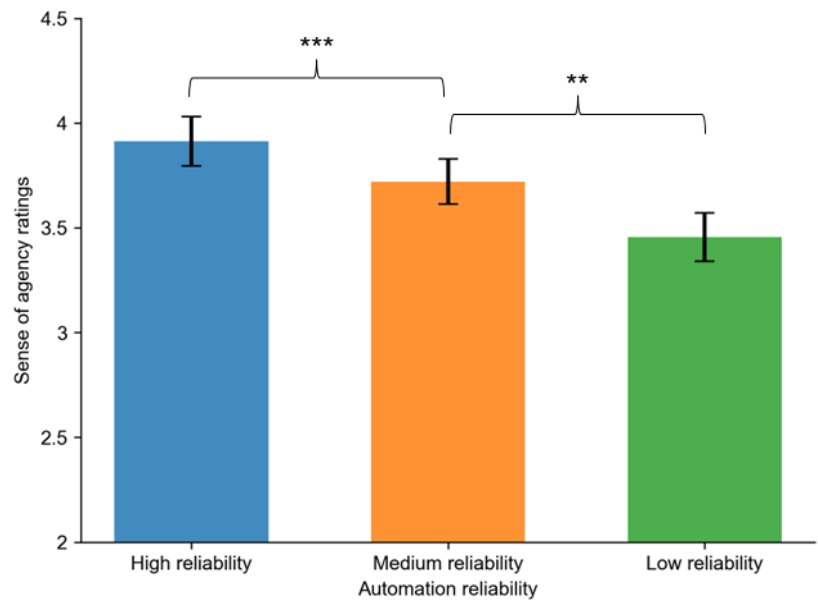

Figure 7

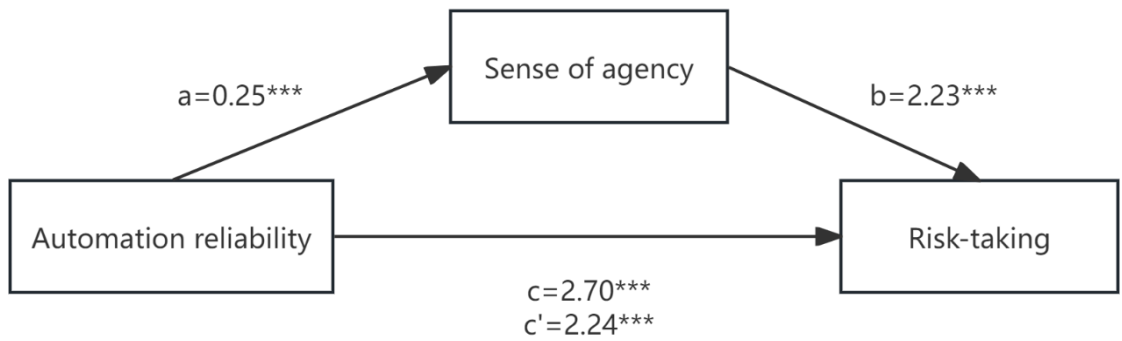

Figure 8

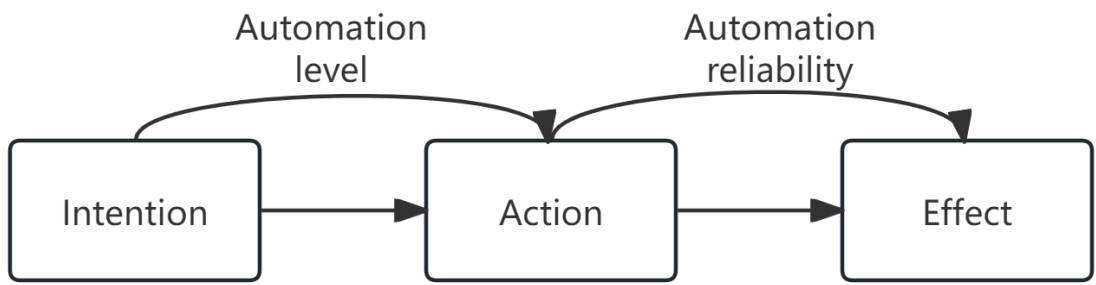

Figure 9

**Figure captions**

Figure 1. Average pumps on unexploded balloons: non-automation vs. automation conditions. Error bars in all figures represent the standard error.

Figure 2 Average pumps on unexploded balloons by automation level

Figure 3 Average sense of agency ratings by automation level

Figure 4 Mediation of sense of agency between automation level and risk-taking

Figure 5 Location and duration of the "Warning" appearance

Figure 6 Average pumps on unexploded balloons by automation reliability

Figure 7 Average sense of agency ratings by automation reliability